Approximate analytical solutions of Dirac Equation with spin and pseudo spin symmetries for the diatomic molecular potentials plus a tensor term with any angular momentum

**Huseyin Akcay** [1,a] **and Ramazan Sever** [2,b]

[1] Faculty of Engineering, Başkent University, Baglıca Campus, Ankara,Turkey

[2] Department of Physics, Faculty of Arts and Sciences, Middle East Technical

University, 06531 Ankara, Turkey

### Abstract

Approximate analytical solutions of the Dirac equation are obtained for some diatomic molecular potentials plus a tensor interaction with spin and pseudospin symmetries with any angular momentum. We find the energy eigenvalue equations in the closed form and the spinor wave functions by using an algebraic method. We also perform numerical calculations for the Pöschl-Teller potential to show the effect of the tensor interaction. Our results are consistent with ones obtained before.

PACS : 03.65.Nk, 03.65.Ge, 03.65.Pm, 03.65.Db, 34.20.-b

Keywords: Diatomic molecular potential; Exact solutions; Bound-states; Energy eigenvalues; Wave functions; Dirac equation; Algebraic method.

[a] E-mail: akcay@baskent.edu.tr

[b] E-mail: sever@metu.edu.tr

# 1. Introduction

To study motion of spin ½ particles in the relativistic approach, Dirac equation is solved to get a complete information about the problems in high energy and nuclear physics. Various techniques has been used in the solution. Some of them are supersymmetry (SUSY) [1], Nikiforov–Uvarov (NU) [2], asymptotic iteration method(AIM) [3-7], factorization and path integral [8-10], shape invariance [11,12]. The pseudospin symmetry is a concept used in nuclear physics to explain the observed degeneracies of certain shell-model orbitals [13-15]. Recently it is shown that [16-18] this symmetry arises from a symmetry of the Dirac Hamiltonian. The Dirac Hamiltonian with external scalar, S(r), and vector, V(r), potentials is invariant for two limits, V-S=constant and V+S=constant. The first one is called the spin symmetry and has applications to the spectrum of mesons and the spectrum of antinucleon [19]. The second limit leads to pseudospin symmetry. This symmetry refers to quasi-degeneracy of the nucleon doublets which can be characterized with quantum numbers $(n, \ell, j = \ell + 1/2)$ and $n-1, \ell+2, j = \ell+3/2)$. Where $n, \ell, j$ are the single nucleon radial, orbital and total angular momentum quantum numbers, respectively. This doublet structure can be expressed in terms of a pseudo-orbital angular momentum $\tilde{\ell} = \ell+1$ and a pseudospin $\tilde{s} = 1/2$. Exact pseudospin symmetry means the degeneracy of the doublets with quantum numbers $j = \tilde{\ell} \pm \tilde{s}$. In recent years, the analytic solutions of the Dirac equation are investigated under the condition of spin and pseudospin symmetries for some typical potentials [20-25]. In addition to scalar and vector potentials a tensor interaction is also considered [20,22,26-29] extensively. Several physical interactions can be described with a tensor potential, such as the interaction of a particle with an anomalous magnetic moment.

In this work we will use an algebraic method to find analytical solutions of the Dirac equation for some well-known diatomic molecular potentials. The same method is used before to study these potentials in the framework of the Schrödinger equation [30]. Diatomic molecular potentials are very important to describe the intramolecular and intermolecular interactions and atomic pair correlations in quantum mechanics. We will study Pöschl-Teller potential [31-38]. Morse Potential [39-41], Mie potential [42-49], Pesudoharmonic potential and Kratzer-Fues Potential [50, 51]

This paper is organized as follows. In section 2, the Dirac equation is solved for the scalar, vector plus a tensor interaction. In section 3, we obtain the solutions of the Dirac equation for the Pöschl-Teller potential plus a Coulomb-like tensor interaction for the spin and pseudospin symmetric cases. In order to understand the effect of the tensor term on the energy spectrum of the bound Dirac states, we list some numerical values in Tables 1 and 2. We show that the Coulomb-like tensor interaction breaks the spin and the pseudospin symmetries but do not change the main characteristics of the energy spectrum**.** We also solve Dirac equation for Morse, Mie, Pesudoharmonic and Kratzer-Fues Potentials. Finally, concluding remarks are given in Sect. 4.

## 2. Dirac equation with scalar, vector and tensor potential

The Dirac equation for a fermion in an external scalar potential S, a vector potential V and a tensor potential U can be written [22, 26, 52] as ($\hbar=1$, c=1)

$$[\vec{\alpha}.\vec{p}+\beta(m+S)+V-i\beta\vec{\alpha}.\hat{r}U]\psi = E\psi \tag{1}$$

Where $\vec{\alpha}$ and $\beta$ are the Dirac matrices. The Hamiltonian

$H_D = \vec{\alpha}.\vec{p}+\beta(m+S)+V-i\beta\vec{\alpha}.\hat{r}U$ commutes with the total angular momentum operator $\vec{J}=\vec{L}+\vec{S}$, with the parity operator and with the spin-orbit operator $K=-\beta(\vec{\sigma}.\vec{L}+I)$. Here $\vec{L}$ is the orbital angular momentum operator and $\vec{s}$ is the spin operator. Hence the eigenfunctions can be written as

$$\psi_{njm\kappa} = \begin{pmatrix} i\dfrac{g_{n\kappa}(r)}{r}\chi^{\ell}_{jm\kappa}(\vartheta,\phi) \\ \dfrac{f_{n\kappa}(r)}{r}\chi^{\tilde{\ell}}_{jm(-\kappa)}(\vartheta,\phi) \end{pmatrix} \tag{2}$$

where $\chi^{\ell}_{jm\kappa}$ denote the spin spherical harmonics. The spin-orbit quantum number [15] $\kappa$ is related to the orbital angular momentum $\ell$ and to the total angular momentum $j$ as follows

$$\kappa = \begin{cases} -(\ell+1) = -(j+\dfrac{1}{2})\ , & j=\ell+\dfrac{1}{2} \quad \text{aligned spin } (\kappa<0) \\ +\ell = +(j+\dfrac{1}{2})\ , & j=\ell-\dfrac{1}{2} \quad \text{unaligned spin } (\kappa>0) \end{cases} \tag{3}$$

In Eq. (2) the lower component of the spinor has a pseudo orbital quantum number $\tilde{\ell}$ which can be expressed as $\tilde{\ell}=\ell-\kappa/|\kappa|$. By substituting the spinor from Eq. (2) into Eq. (1) we obtain the following set of coupled equations [52] for the radial components $g_{n\kappa}$ and $f_{n\kappa}$.

$$(\dfrac{d}{dr}+\dfrac{\kappa}{r}-U(r))g_{n\kappa}(r) = (E+m-\Delta)f_{n\kappa} \tag{4a}$$

$$(\dfrac{d}{dr}-\dfrac{\kappa}{r}+U(r))f_{n\kappa}(r) = -(E-m-\Sigma)g_{n\kappa} \tag{4b}$$

where $\Delta = V - S$ and $\Sigma = V + S$. We note that in Ref. (21) the Dirac equation is written as $H_D \psi = \varepsilon \psi$ and the parameter $E$ is defined as $E = \varepsilon - m$. Therefore, the parameter $E$, in their equations, is not the eigenvalue of the Dirac Hamiltonian. In our formalism $E$ is the eigenvalue of the Hamiltonian. In order to make a comparison one must replace $E$ with $E + m$ in our equations. By combining Eq. (4a) and Eq. (4b) we get the following second order differential equations for the radial wave functions.

$$(\frac{d^2}{dr^2} - \frac{\kappa(\kappa+1)}{r^2} + \frac{2\kappa}{r}U - \frac{dU}{dr} - U^2)g_{n\kappa}(r) + \frac{\frac{d\Delta}{dr}}{M_\Delta}(\frac{d}{dr} + \frac{\kappa}{r} - U)g_{n\kappa}(r) \quad (5a)$$
$$= -(M_\Delta M_\Sigma)g_{n\kappa}(r)$$

$$(\frac{d^2}{dr^2} - \frac{\kappa(\kappa-1)}{r^2} + \frac{2\kappa}{r}U + \frac{dU}{dr} - U^2)f_{n\kappa}(r) + \frac{\frac{d\Sigma}{dr}}{M_\Sigma}(\frac{d}{dr} - \frac{\kappa}{r} + U)f_{n\kappa}(r) \quad (5b)$$
$$= -(M_\Delta M_\Sigma)f_{n\kappa}(r).$$

where $M_\Delta = (E + m - \Delta)$ and $M_\Sigma = (E - m - \Sigma)$.

As stated before, we use an algebraic method for the solution of these radial wave equations. In this scheme the wave equation is transformed into a second order parametric differential equation whose solution are studied before. In the following section we apply this method for the solution of Eq. (5) for several molecular potentials.

### 3. Applications

### a) Pöschl-Teller potential

We start to study the bound state solution for this potential under the assumption of spin symmetry case. Hence we assume that $\Delta$ is given by a constant C and the sum of the potentials is given by the Pöschl-Teller [31-38] potential. That is

$$\Sigma = -\frac{A(A+\alpha)}{\cosh^2(\alpha.r)} + \frac{B(B-\alpha)}{\sinh^2(\alpha.r)} \quad (9)$$

where the parameter $\alpha$ is related to the range of the potential. For the Coulomb-like tensor potential we use the parameterization $U = -\gamma/r$. We note that if we take $\gamma = 0$ at any step in our calculations we obtain the solutions of the Dirac equation without the tensor interaction. So it can be taken as an investigation with scalar and vector potential only. In order to study Eqs. (5a) and (5b) for arbitrary spin-orbit quantum number $\kappa$ the centrifugal term is approximated as [53, 54]

$$\frac{1}{r^2} \approx \frac{4\alpha^2 \exp(-2\alpha r)}{(1-\exp(-2\alpha r))^2} \quad . \tag{10}$$

where $\alpha^2$ is assumed to be small. With this approximation Eq. (5a) can be written in the following form

$$\{\frac{d^2}{dr^2} - \alpha^2 \frac{\ell(\ell+1)}{\sinh^2(\alpha r)} - (\varepsilon + m - C)(-\frac{A(A+\alpha)}{\cosh^2(\alpha r)} + \frac{B(B-\alpha)}{\sinh^2(\alpha r)})$$
$$- \frac{\alpha^2[\gamma(\gamma+1) + 2\kappa\gamma]}{\sinh^2(\alpha r)}\}g_{n\kappa} = -(E-m)(E+m-C)g_{n\kappa} \tag{11}$$

where we have used the identity $\kappa(\kappa+1) = \ell(\ell+1)$ and we have replaced U by $(-\gamma/r)$. This equation takes a simpler form when it is written in terms of a dimensionless variable s defined as $s = \sinh^2(\alpha r)$. Using this new variable we transform Eq. (11) into the following equation

$$\frac{d^2 g_{nk}}{ds^2} + \frac{(s+\frac{1}{2})}{s(s+1)} \frac{dg_{n\kappa}}{ds} + \frac{H_1 s^2 + (H_1 + L_1 - F_1)s - F_1}{[s(s+1)]^2} g_{n\kappa} = 0 \tag{12}$$

where

$$H_1 = \frac{(E-m)(E+m-C)}{4\alpha^2} \tag{13a}$$

$$L_1 = \frac{(E+m-C)A(A+\alpha)}{4\alpha^2} \tag{13b}$$

$$F_1 = \frac{(E+m-C)B(B-\alpha)}{4\alpha^2} + \frac{1}{4}[\ell(\ell+1) + \gamma(\gamma+1) + 2\kappa\gamma] \tag{13c}$$

Eq. (12) has the form of the parametric differential equation that was mentioned before. The general form [30] is given as

$$\frac{d^2\varphi}{ds^2} + \frac{(c_1 + c_2 s)}{s(1+c_3 s)} \frac{d\varphi}{ds} + \frac{1}{s^2(1+c_3 s)^2}[-\Lambda_1 s^2 + \Lambda_2 s - \Lambda_3]\varphi = 0 \tag{14}$$

Where $c_i$ and $\Lambda_i$ are some constants. In the following we use the results and the notation of Ref. (30). When we compare Eq. (12) with Eq. (14) we find that they are the same type. The parameters can be expressed as

$$c_1 = \frac{1}{2}, \quad c_2 = 1, \quad c_3 = 1, \quad \Lambda_1 = -H_1, \quad \Lambda_2 = (H_1 + L_1 - F_1), \quad \Lambda_3 = F_1. \tag{15}$$

The bound state solutions of Eq. (14) are given in detail in Ref. (30). Depending on the value of the parameter $c_3$ there are two possibilities. When the parameter $c_3$ is not zero the solutions are

$$\varphi(s) = (1+c_3 s)^{-p_0} s^{q_0} P_n^{(\alpha,\beta)}(1+2c_3 s) \tag{16}$$

where

$$q_0 = (\frac{1-c_1}{2}) \pm \sqrt{(\frac{1-c_1}{2})^2 + \Lambda_3}, \qquad p_0 = \frac{D}{2} \pm \sqrt{(\frac{D}{2})^2 + H} \tag{17}$$

$$D = \frac{c_2}{c_3} - c_1 - 1, \qquad H = \frac{\Lambda_1}{c_3^2} + \frac{\Lambda_2}{c_3} + \Lambda_3 \tag{18}$$

Here the functions $P_n^{(\alpha,\beta)}(z)$ are the Jacobi polynomials and $\alpha = 2q_0 + c_1 - 1$, $\beta = -2p_0 - c_1 + \frac{c_2}{c_3} - 1$. The formula for the corresponding energy levels is given as

$$(q_0 - p_0)^2 + (\frac{c_2}{c_3} + 2n - 1)(q_0 - p_0) + n(n + \frac{c_2}{c_3} - 1) = \frac{\Lambda_1}{c_3^2} \tag{19}$$

As stated before, these formulas are derived in Ref. (30) Inserting $c_i$'s into this equation it can written as

$$[(q_0 - p_0) + n]^2 = \Lambda_1 \tag{20}$$

and using Eqs. (15) and (13) $q_0, p_0, \alpha$ and $\beta$ can be written, in terms of $H_1, L_1, F_1$ as

$$q_0 = \frac{1}{4} + \frac{1}{4}\sqrt{1+16F_1}, \qquad p_0 = -\frac{1}{4} + \frac{1}{4}\sqrt{1+16L_1}, \tag{21}$$

$$\alpha = 2q_0 - \frac{1}{2}, \qquad \beta = -(2p_0 + \frac{1}{2}) \tag{22}$$

The parameters in these equations are defined in Eq. (13). Inserting all these into Eq. (20) and using Eq. (13) we obtain

$$(E-m)(E+m-C) = -4\alpha^2 \{ \frac{2n+1}{2} - \frac{1}{4}[1+\frac{4}{\alpha^2}(E+m-C)A(A+\alpha)]^{1/2}$$
$$+ \frac{1}{4}[1+\frac{4}{\alpha^2}(E+m-C)B(B-\alpha) + 4[\ell(\ell+1) + \gamma(\gamma+1) + 2\kappa\gamma]^{1/2} \}^2 \tag{23}$$

The corresponding radial wave functions are given by

$$g_{n\kappa}(s) = (1+s)^{-p_0} s^{q_0} P_n^{(\alpha,\beta)}(1+2s)$$

It is known that there are no bound negative energy states in the limit of spin symmetry and no bound positive energy states in the pseudospin symmetry limit [15,19]. The influence of tensor interactions on these spectra has been also investigated [20, 21, 22, 26]. For instance, tensor potential U linear in r is investigated and its contributions to the positive and negative energy solutions of the Dirac equation are studied explicitly in Ref. (20). But here we have a Coulomb-like tensor potential and in order to understand the effect of this potential on the spectrum of the bound states we take a set of values for our parameters [56, 57] and calculate energy eigenvalues. Using Eq. (23) we obtain the values listed in Table 1. We can see the splitting of the energy levels of the spin doublets . Fig. 1 indicates the dependence of this splitting on the tensor interaction strength. With these parameters we found that Eq. (23) has only positive energy solutions for bound states. That is, the added tensor term does not lead to negative energy solutions. We can understand this as follows. The tensor potential enters into the wave equation as $(\Omega/r^2)$ where $\Omega = \kappa(\kappa+1) + 2\kappa\gamma + \gamma(\gamma+1)$. Table 2 shows how the parameter $\Omega$ is changing with the coupling constant $\gamma$. We can see that $\Omega$ remains positive for all values of $\gamma$ and thus $\Omega/r^2$ acts as centrifugal barrier. Thus the binding is the result of the other potentials present in the problem. We also note that this is valid for all the cases we are studying in this work.

Pseudospin symmetry case:

In order to study the solutions for the pseudospin symmetry case we must choose $\Sigma$ as a constant. For $\Delta$ we assume

$$\Delta = -\frac{A(A+\alpha)}{\cosh^2(\alpha.r)} + \frac{B(B-\alpha)}{\sinh^2(\alpha.r)}, \qquad \Sigma = C. \tag{24}$$

We take the tensor potential with the same parameterization and assume the same approximation for the centrifugal term. After these assumptions we write Eq. (5b) in terms of the variable s. The result is the following equation

$$\frac{d^2 f_{n\kappa}}{ds^2} + \frac{(s+\frac{1}{2})}{s(s+1)} \frac{df_{n\kappa}}{ds} + \frac{H_2 s^2 + (H_2 + L_2 - F_2)s - F_2}{[s(s+1)]^2} f_{n\kappa} = 0 \tag{25}$$

where

$$H_2 = \frac{(E+m)(E-m-C)}{4\alpha^2} \tag{26a}$$

$$L_2 = \frac{(E-m-C)A(A+\alpha)}{4\alpha^2} \tag{26b}$$

$$F_2 = \frac{(E-m-C)B(B-\alpha)}{4\alpha^2} + \frac{1}{4}[\tilde{\ell}(\tilde{\ell}+1) + \gamma(\gamma-1) + 2\kappa\gamma] . \tag{26c}$$

We observe that Eq. (25) has the same form as Eq. (12), only the parameters are different. Therefore we can write down the solutions for Eq. (25) by replacing $H_1, L_1, F_1$ with $H_2, L_2, F_2$ in the solutions of Eq. (12). This way we obtain the following solutions

$$f_{n\kappa}(s) = s^{q_0}(1+s)^{-p_0} P_n^{(2q_0-1/2, 2p_0-1/2)}(1+2s) . \tag{27}$$

where

$$q_0 = \frac{1}{2}(\frac{1}{2}\sqrt{1+16F_2} + \frac{1}{2}) , \quad p_0 = \frac{1}{2}(\frac{1}{2}\sqrt{1+16L_2} - \frac{1}{2}) , \tag{28}$$

To find the energy eigenvalues we follow the steps of the previous sections. This leads to the following formula for the energy spectrum

$$(E+m)(E-m-C) = -4\alpha^2 \{ \frac{2n+1}{2} - \frac{1}{4}[1 + \frac{4}{\alpha^2}(E-m-C)A(A+\alpha)]^{1/2}$$
$$+ \frac{1}{4}[1 + \frac{4}{\alpha^2}(E-m-C)B(B-\alpha) + 4[\tilde{\ell}(\tilde{\ell}+1) + \gamma(\gamma-1) + 2\kappa\gamma]^{1/2} \}^2 \tag{29}$$

We see that this equation is not identical with Eq. (23). So it gives a different spectrum. If we use the same numerical values for parameters given in the previous section we get only negative energy bound state solutions. This is the case for the exact pseudospin symmetry and this can be explained as in the previous section.

**b) Morse potential**

We will study the bound state solutions with the following potentials [39-41]

$$\Sigma = D[\exp(-2\beta(r-r_0)) - 2\exp(-\beta(r-r_0))] \tag{30a}$$

$$U = \frac{\gamma}{r}, \qquad \Delta = C \tag{30b}$$

where α and C are constants. Hence this gives a spin symmetric Hamiltonian. D corresponds to the depth of the Morse potential, β is related to the range of the potential and $r_0$ is the equilibrium internuclear distance. Inserting Eq. (30) into Eq. (5a) we find

$$\begin{aligned}\{\frac{d^2}{dr^2} - \frac{\kappa(\kappa+1)}{r^2} - (E+m-C)D[\exp(-2\beta(r-r_0)) - 2\exp(-\beta(r-r_0))] \\ - \frac{\gamma(\gamma+1)+2\kappa\gamma}{r^2}\} g_{n\kappa} = -(E-m)(E+m-C)g_{n\kappa}\end{aligned} \tag{31}$$

Let us define a new variable $\rho = (r-r_0)/r_0$ and write Eq. (31) in terms of this new variable. We find

$$\begin{aligned}\{\frac{d^2}{d\rho^2} - \frac{B}{r_0^2(1+\rho)^2} - (E+m-C)D[\exp(-2\alpha\rho) - 2\exp(-\alpha\rho)]\}g_{n\kappa} \\ = -(E-m)(E+m-C)g_{n\kappa}\end{aligned} \tag{32}$$

where $\alpha = r_0\beta$ and $B = \kappa(\kappa+1) + 2\kappa\gamma + \gamma(\gamma+1)$. In the following, we use the identity $\kappa(\kappa+1) = \ell(\ell+1)$. For the second term we are going to use the Pekeris approximation [54, 55]. Within this approximation it can be written as

$$\frac{1}{(1+\rho)^2} \cong \{D_0 + D_1 \exp(-\alpha\rho) + D_2 \exp(-2\alpha\rho)\} \tag{33}$$

where

$$D_0 = 1 - \frac{3}{\alpha} + \frac{3}{\alpha^2}, \quad D_1 = \frac{4}{\alpha} - \frac{6}{\alpha^2}, \quad D_2 = -\frac{1}{\alpha} + \frac{3}{\alpha^2} \;.$$

Since the expression $\exp(-\alpha\rho)$ occurs many times, we define a new dimensionless variable s as $s = \exp(-\alpha\rho)$. In terms of this new variable Eq. (32) transforms into the following equation.

$$\frac{d^2 g_{n\kappa}}{ds^2} + \frac{1}{s}\frac{dg_{n\kappa}}{ds} + \frac{1}{s^2}[-\Lambda_1 s^2 + \Lambda_2 s - \Lambda_3]g_{n\kappa} = 0 \tag{34a}$$

where

$$\Lambda_1 = \frac{BD_2}{\alpha^2} + \frac{(E+m-C)Dr_0^2}{\alpha^2}, \quad \Lambda_2 = -\frac{BD_1}{\alpha^2} + 2\frac{(E+m-c)Dr_0^2}{\alpha^2},$$

$$\Lambda_3 = \frac{BD_0}{\alpha^2} - \frac{(E-m)(E+m-C)r_0^2}{\alpha^2} \tag{34b}$$

To find the radial wave functions $g_{n\kappa}$, we must solve Eq. (34a). This equation has the form given in Eq. (14) and its parameters are $c_1 = 1$, $c_2 = 0$, $c_3 = 0$. As discussed before [30], when $c_3$ is zero the bound state solutions of Eq. (14) are given by

$$\varphi(s) = \exp(-p_{10}s) s^{q_{10}} L_n^k[(2p_{10} - c_2)s] \tag{35}$$

where

$$q_{10} = \frac{(1-c_1)}{2} + \sqrt{(\frac{1-c_1}{2})^2 + \Lambda_3}, \quad p_{10} = \frac{c_2}{2} + \sqrt{(\frac{c_2}{2})^2 + \Lambda_1}, \quad k = c_1 + 2q_{10} - 1 \tag{36}$$

and the energy levels are given as

$$c_1 p_{10} - q_{10}(c_2 - 2p_{10}) - \Lambda_2 = n(c_2 - 2p_{10}) \tag{37}$$

Inserting $c_i$'s into Eq. (36) we find $q_{10} = \sqrt{\Lambda_3}$, $p_{10} = \sqrt{\Lambda_1}$ and Eq. (37) gives

$$(2n+1+2\sqrt{\Lambda_3})\sqrt{\Lambda_1} = \Lambda_2. \tag{38}$$

We solve this equation for $\Lambda_3$ and obtain

$$\Lambda_3 = \{\frac{\Lambda_2}{2\sqrt{\Lambda_1}} - (n+\frac{1}{2})\}^2. \tag{39}$$

Finally replacing $\Lambda_1$, $\Lambda_2$, $\Lambda_3$ with the expressions given in Eq. (34b) we find

$$(E-m)(E+m-C) = \frac{BD}{r_0^2} - \frac{1}{r_0^2}\{\frac{(2Dr_0^2(E+m-C) - BD_1)}{2\sqrt{Dr_0^2(E+m-C) + BD_2}} - (n+\frac{1}{2})\gamma\}^2 \tag{40}$$

Thus the energy levels depend on the quantum numbers n, $\ell$ and $\kappa$. The corresponding wave function are

$$g_{n\kappa}(s) = \exp(-\sqrt{\Lambda_1}\, s)\, s^{\sqrt{\Lambda_3}} L_n^{2\sqrt{\Lambda_3}}(2\sqrt{\Lambda_1}\, s) \qquad (41)$$

## Pseudospin symmetry case

For pseudospin symmetry we must choose $\Sigma$ as constant. This time $\Delta$ is assumed to be given by

$$\Delta = D[\exp(-2\beta(r-r_0)) - 2\exp(-\beta(r-r_0))]. \qquad (42)$$

We take the tensor potential with the same parameterization and assume the same approximation for the centrifugal term. After these assumptions we write Eq. (5b) in terms of the variable s. The result is the following equation

$$\frac{d^2 f_{n\kappa}}{ds^2} + \frac{1}{s}\frac{df_{n\kappa}}{ds} + \frac{1}{s^2}[-\Lambda_{21} s^2 + \Lambda_{22} s - \Lambda_{23}] f_{n\kappa} = 0 \qquad (43a)$$

where

$$\Lambda_{21} = \frac{BD_2}{\alpha^2} + \frac{(E-m-C)Dr_0^2}{\alpha^2}, \quad \Lambda_{22} = -\frac{BD_1}{\alpha^2} + 2\frac{(E-m-c)Dr_0^2}{\alpha^2},$$

$$\Lambda_{23} = \frac{BD_0}{\alpha^2} - \frac{(E+m)(E-m-C)r_0^2}{\alpha^2} \qquad (43b)$$

Eq. (43a) has the same form as Eq. (34a), only the parameters are different. Hence we can write down the solutions for Eq. (43a) by replacing $\Lambda_1, \Lambda_2, \Lambda_3$ with $\Lambda_{21}, \Lambda_{22}, \Lambda_{23}$ in equations (36, 37, 38). The result for the wave function is

$$f(s) = \exp(-\sqrt{\Lambda_{21}}\, s)\, s^{\sqrt{\Lambda_{23}}} L_n^{2\sqrt{\Lambda_{23}}}(2\sqrt{\Lambda_{21}}\, s). \qquad (44)$$

And the equation for the corresponding energy levels is

$$(E+m)(E-m-C) = \frac{BD}{r_0^2} - \frac{1}{r_0^2}\left\{\frac{(2Dr_0^2(E-m-C) - BD_1}{2\sqrt{Dr_0^2(E-m-C) + BD_2}} - (n+\frac{1}{2})\gamma\right\}^2 \qquad (45)$$

## c) Mie Potential

To apply our formulation to the Dirac equation for the Mie Potential [42-49] we again start from Eq. (5a). Assume that

$$\Sigma = V_0[\frac{1}{2}(\frac{a}{r})^2 - \frac{a}{r}] \quad , \quad \Delta = C \tag{46}$$

and take the tensor potential with the same parameterization. Inserting these into Eq. (5a) we find

$$(\frac{d^2}{dr^2} - \frac{\kappa(\kappa+1) + \gamma(\gamma+1) + 2\kappa\gamma}{r^2}) g_{n\kappa}(r) = -(E+m-C)(E-m-\Sigma) g_{n\kappa}(r) \tag{47}$$

Let us define $G = g/r$ and $s = r$ and express this equation for G in terms of the variable s. We obtain

$$\frac{d^2 G}{ds^2} + \frac{2}{s}\frac{dG}{ds} + \frac{1}{s^2}[-\Lambda_1 s^2 + \Lambda_2 s - \Lambda_3]G = 0 \tag{48}$$

where

$$\Lambda_1 = -(E+m-C)(E-m), \quad \Lambda_2 = aV_0(E+m-C),$$

$$\Lambda_3 = l(l+1) + \gamma(\gamma+1) + 2\kappa\gamma + a^2\frac{V_0}{2}(E+m-C) \tag{49}$$

When we compare Eq. (48) with Eq. (14) we find that $c_1 = 2$, $c_2 = 0$, $c_3 = 0$. Thus the solutions are given by Eqs. (35, 36, 37). Inserting the parameters into these equations we find

$$q_{10} = \frac{1}{2}(-1 + \sqrt{1+4\Lambda_3}), \quad p_{10} = \sqrt{\Lambda_1}, \quad k = \sqrt{1+4\Lambda_3} . \tag{50}$$

Hence the wave functions are

$$g_{n\kappa}(s) = \exp(-\sqrt{\Lambda_1} s) s^{\frac{1}{2}(\sqrt{1+4\Lambda_3}+1)} L_n^{(\sqrt{1+4\Lambda_3})}(2\sqrt{\Lambda_1} s) \tag{51}$$

and the corresponding energy values are given as

$$(E+m-C)(E-m) = -\Lambda_2^2 [2n+1+\sqrt{1+4\Lambda_3}]^{-2} \tag{52}$$

If we take $C = 0$, $\gamma = 0$ and go to the none relativistic limit by replacing $E+m$ with 2m and $E-m$ with $\varepsilon$ we obtain

$$\varepsilon_n = -\frac{1}{2m}[2n+1+\sqrt{1+4[\frac{2ma^2 V_0}{2} + l(l+1)]}](2maV_0)^2 . \tag{53}$$

This spectrum agrees with the results obtained in Ref. (30) for the Schrödinger equation with the Mie Potential. It is easy to obtain the pseudospin symmetric solutions for this potential. As it is demonstrated for the two previous examples this can be done easily. So we are not going to give the solutions for the pseudosin symmetry case for Mie Potential and for the following applications.

**d) Pseudoharmonic Potential**

The pseudoharmonic potential is given as $V(r) = V_0(\frac{r}{r_0} - \frac{r_0}{r})^2$ and we will choose $\Delta = C$ and $\Sigma = V(r)$. The tensor potential will be taken as before. When we replace these into Eq. (5a) and define a new variable $s = r^2$ and a new function $G = g/r$ we end up with the following equation.

$$\frac{d^2 G}{ds^2} + \frac{3}{2}\frac{1}{s}\frac{dG}{ds} + \frac{1}{s^2}\{-\Lambda_1 s^2 + \Lambda_2 s - \Lambda_3\}G = 0 \tag{54}$$

where

$$\Lambda_1 = \frac{1}{4r_0^2}(E+m-C)V_0$$

$$\Lambda_2 = \frac{(E+m-C)}{4}[2V_0 + (E-m)] \tag{55}$$

$$\Lambda_3 = \frac{1}{4}[l(l+1) + \gamma(\gamma+1) + 2\kappa\gamma + (E+m-C)V_0 r_0^2]$$

Eq. (55) gives $c_1 = 3/2$, $c_2 = 0$, $c_3 = 0$ and using these in Eq. (36) obtain

$$q_{10} = \frac{1}{4}(\sqrt{1+16\Lambda_3} - 1), \quad p_{10} = \sqrt{\Lambda_1}, \quad k = \frac{1}{2}\sqrt{1+16\Lambda_3} \tag{56}$$

Thus the wave functions can be written as

$$g_{n\kappa}(s) = \exp(-\sqrt{\Lambda_1} s) s^{\frac{(\sqrt{1+16\Lambda_3}+2)}{4}} L_n^{\frac{\sqrt{1+16\Lambda_3}}{2}}(2\sqrt{\Lambda_1} s) \tag{57}$$

and the corresponding energy values can be obtained from Eq.( 37). We get

$$\sqrt{\Lambda_1}[2n+1+\frac{1}{2}\sqrt{1+16\Lambda_3}] = \Lambda_2 \tag{58}$$

or replacing $\Lambda_1$ and $\Lambda_2$ we obtain

$$(E+m-C)(2V_0 + E - m) = 4\{\sqrt{\frac{(E+m-C)}{4r_0^2}}[2n+1+\frac{1}{2}\sqrt{1+16\Lambda_3}]\} \tag{59}$$

When this equation is written in the nonrelativistic limit with $C = 0$, $\gamma = 0$ the result is

$$(2V_0 + \varepsilon) = \frac{4}{2m}\{2n + 1 + \frac{1}{2}\sqrt{1 + 4[l(l+1) + 2mV_0 r_0^2]}\}\sqrt{\frac{2mV_0}{4r_0^2}} \quad . \tag{60}$$

This is the result obtained in Ref. (30) for the Schrödinger equation for this potential.

### e) Kratzer-Fues Potential

The Kratzer-Fues [50, 51] potential is given by

$$V(r) = D_e(\frac{r - r_e}{r})^2 \quad .$$

We chose $\Delta = C$, $\Sigma = V(r)$ and define $G = g/r$, s=r and inserting these into Eq. (5a). The result is the following equation

$$\frac{d^2G}{ds^2} + \frac{2}{s}\frac{dG}{ds} + \frac{1}{s^2}[-\Lambda_1 s^2 + \Lambda_2 s - \Lambda_3]G = 0 \tag{62}$$

where

$$\Lambda_1 = (E + m - C)D_e - (E + m - C)(E - m)$$

$$\Lambda_2 = 2r_e(E + m - C)D_e$$

$$\Lambda_3 = l(l+1) + \gamma(\gamma + 1!) + 2\kappa\gamma + r_e^2(E + m - C)D_e \quad .$$

Using Eq. (62) we write the parameters as $c_1 = 2$, $c_2 = 0$, $c_3 = 0$ and thus

$$q_{10} = \frac{1}{2}(\sqrt{1 + 4\Lambda_3} - 1), \quad p_{10} = \sqrt{\Lambda_1}, k = 1 + \sqrt{1 + 4\Lambda_3} \quad . \tag{63}$$

Thus the solutions are

$$g_{n\kappa} = \exp(-\sqrt{\Lambda_1} s)s^{\frac{1}{2}(\sqrt{1+4\Lambda_3}+1)} L_n^{(1+\sqrt{1+4\Lambda_3})}(2\sqrt{\Lambda_1} s) \tag{64}$$

and the energy levels satisfy the following equation

$$(E - m - D_e) = -\frac{4r_e^2(E + m - C)}{[2n + 1 + \sqrt{1 + 4\Lambda_3}]^2} \quad . \tag{65}$$

Its nonrelativistic limit with $C = 0, \gamma = 0$ is

$$\varepsilon - D_e = -\frac{4r_e^2 2m}{[2n+1+\sqrt{1+4[l(l+1)+2mr_e^2 D_e]}]^2} \tag{66}$$

and this also agrees with the Schrödinger equation.

## 3. Conclusions

We have obtained the approximate analytical solutions of the Dirac equation with spin and pseudospin symmetry for some well-known diatomic potentials plus a Coulomb like tensor interaction. An algebraic method is used in the calculations to obtain energy eigenvalues in the closed form and the corresponding spinor wave functions. We have listed the numerical results of the energy eigenvalues for the spin symmetric case inTables 1. The variation of the parameter
$\Omega(=\kappa(\kappa+1)+2\kappa\gamma+\gamma(\gamma+1))$ as a function of the coupling constant $\gamma$ for two states is presented in Tanle 2. We have shown that the tensor interaction removes the degeneracy between the members of doublet states.

### 4. ACKNOWLEDGMENTS

This research was partially supported by the Scientific and Technical Research Council of Turkey.

References

[1] G. Levai: On some exactly solvable potentials derived from supersymmetric quantum -mechanics. J. Phys. A: Math. Gen. 25, L521(1992).

[2] Nikiforov A F and Uvarov V B :1988 Special Functions of Mathematical Physics (New York: Academic).

[3] H. Ciftci, R. L. Hall and N. Saad: Asymptotic iteration method for eigenvalue problems . J. Phys. A: Math. Gen. 36, 11807(2003).

[4] H. Ciftci, R. L. Hall and N. Saad: Construction of exact solutıons to eigenvalue problems by the asymptotic iteration method. J. Phys. A: Math. Gen. 38, 1147(2005).

[5] O. Bayrak, I. Boztosun: Arbitrary l-state solutions of the rotating Morse potential by the asymptotic iteration method. J. Phys. A:Math. Gen. 39, 6955 (2006).

[6] F. Yasuk , A. Durmus and I Boztosun: Exact analytical solution to the relativistic Klein-Gordon equation with noncentral equal scalar and vector potentials. J. Math. Phys. 47, 082302 (2006).

[7] I. Boztosun, M. Karakoc, F. Yasuk and A. Durmus: Asymptotic iteration method solutions to the relativistic Duffin-Kemmer-Petiau equation. J. Math Phys. 47, 062301(2006).

[8] L. Infeld and T. E. Hull: The Factorization method. Rev. Mod. Phys. 23, 21(1951).

[9] A. Stahlhofen: An algebraic form of the factorization method. Nuovo Cimento B 104, 447(1989).

[10] R. M. Edelstein, K. S. Govinder and F. M. Mahomed : Solution of ordinary differential equations via nonlocal transformations. J. Phys. A: Math. Gen. 34, 1141(2001).

[11] L. Gendenshtein, Pisma Zh. Eksp. Teor. Fiz. Piz. Red. 38 (1983) 299 (JETP Lett. 38 (1983) 356).

[12] L. Gendenshtein: Derivation of exact spectra of the schrodinger-equation by means of supersymmetry. JETP Lett. 38, 356(1983).

[13] K. T. Hecht and A Adler: Generalized seniority for favored $J \neq 0$ pairs in mixed configurations , Nucl. Phys. A 137, 129 (1969).

[14] A. Arima , M. Harvey and K Shimizu: Pseudo LS coupling and pesudo SU3 copuling schemes. Phys. Lett. B 30, 517. (1969).

[15] J. N. Ginocchio: Relativistic symmetries in nuclei and hadrons. Phys. Rep.414,165 (2005).

[16] J. N. Ginocchio: Pseudospin as a relativistic symmetry. Phys. Rev. Lett. 78, 437 (1997).

[17] P. R. Page , T Goldman and J N Ginocchio: Relativistic symmetry suppresses quark spin-orbit splitting. Phys. Rev. Lett.86, 204(2001).

[18] J. N. Ginocchio, A Leviatan, J Meng, and S G Zhou: Test of pseudospin symmetry in deformed nuclei. Phys.Rev. C 69, 034303(2004).

[19] J. N. Ginocchio: Relativistic harmonic oscillator with spin symmetry. Phys. Rev. C 69, 034318 (2004).

[20] R. Lisboa, M Malheiro, A S de Castro, P Alberto and M Fiolhais: Pseudospin symmetry and the relativistic harmonic oscillator . Phys. Rev. C 69, 024319 (2004)

[21] P. Alberto, R. Lisboa, M. Malheiro, A.S. de Catro: Tensor coupling and pseudospin symmetry in nuclei . Phys. Rev. C 71,034313 (2005)

[22] H. Akcay: Dirac equation with scalar and vector quadratic potentials and Coulomb-like tensor potential. Phys. Lett. A 373, 616 (2009).

[23] J. Y. Guo,Z Q Sheng: Solution of the Dirac equation for the Woods-Saxon potential with spin and pseudospin symmetry. Phys. Lett. A 338, 90 (2005).

[24] C. Berkdemir: Pseudospin symmetry in the relativistic Morse potential including the spin-orbit coupling term. Nucl. Phys. A 770, 32 (2006).

[25] C. S. Jia, P Guo, X L Peng: Exact solution of the Dirac-Eckart problem with spin and pseudospin symmetry  J. Phys. A Math. Theor. 39, 7737(2006).

[26] O. Aydogdu, R. Sever: Pseudospin and spin symmetry in Dirac-Morse problem with a tensor potential  Phys. Lett. B 703, 379 (2011).

[27] O. Aydogdu and R Sever: Exact Pseudospin Symmetric Solution of the Dirac Equation for Pseudoharmonic Potential in the Presence of Tensor Potential. Few-Body Syst. 47, 193 (2010)

[28] M. Hamzavi, A. A. Rajabi, and H. Hassanabadi: Exact Spin and Pseudospin Symmetry Solutions of the Dirac Equation for Mie-Type Potential Including a Coulomb-like Tensor Potential. 2010 Few-Body Syst. 48, 171(2010)

[29] M. Hamzavi, A A Rajabi, and H Hassanabadi: Relativistic Morse Potential and Tensor interaction. Few-Body Syst. 52, 19 (2012)

[30] H. Akcay, R. Sever: Analytical solutions of Schrodinger equation for the diatomic molecular potentials with any angular momentum. J Math Chem. 50,1973 (2012).

[31] G. Pöschl, E. Teller: Bemerkungen sur Quantenmechanik des anharmonishen Oszillators. Z. Phys. 83, 143 (1933).

[32] S. H. Dong, W. C. Qiang, J. G. Ravelo: Analytical approximations to the Schrodinger equation for a second pschlteller-like potential with centrifugal term. Int. J. Mod. Phys. A 23, 1537(2008).

[33] R. Dutt, A. Khare, Y. P. Varshni: New class of conditionally exactly solvable potentials in quantum mechanics. J. Phys. A Math. Gen. 28, L107 (2004).

[34] C. S. Jia, Y Li: Coplexified Poschl-Teller potential model. Phys. Lett. A 305, 231 (2002).

[35] M. Simsek, Z. Yalcin: Generalized Poschl-Teller Potential. J. Math. Chem. 16, 211 (1994).

[36] M. G. Miranda, G. H. Sun, S. H. Dong: The solution of the second poschl-teller like potential by nikiforov-uvarov method. Int. J. Mod. Phys. E-Nucl. Phys. 19, 123(2010).

[37] A. D. Alhaidari: Relativistic extension of shape invariant potential. J. Phys. A:Math. Gen. 34, 9827 (2006).

[38] A. D. Alhaidari, H. Bahlouli, A. Al-Hasan: Dirac and Klein-Gordon equations with equal scalar and vector potentials Phys. Lett. A 349, 87 (2006).

[39] P. M. Morse: Diatomic molecules according to wave mechanics II. Phys. Rev. 34, 57 (1929).

[40] L. H. Zhang, X P Li, C. S. Jia: Approximate Solutions of the Schrodinger Equation with Generalized Morse Potential Model Including the Centrifugal Term, Int. J. Q. Chem. 111, 1870(2011).

[41] N. Saiki, S. A. S. Ahmed: Exact analytic solutions generated from stipulated Morse and trigonometric Scarf potentials. Phys. Scr. 83, 035006(2011).

[42] G. Mie: Zur kinetischen Theorie der einatomigen Körper. Ann. Phys. Leipz. II, 657 (1903).

[43] Y. Weissman, J. Jortner: Isotonic oscillator. Phys. Lett. A 70, 177 (1979).

[44] M. L. Szego: The vibrations of the pseudogaussian oscillator. Chem. Phys. 87, 431 (1984).

[45] M. Sage, J. Goodisman: Improving on the conventional presentatıon of molecular vibratıons - advantages of the pseudoharmonıc potentıal and the constructıon of potentıal-energy curves. Am. J. Phys. 53, 350 (1985).

[46] S. Erkoç, R. Sever: Path-integral solutıon for pseudoharmonic potentıal . Phys. Rev. A 37, 2687 (1988).

[47] R. Sever, M. Bucurgat, C. Tezcan, O. Yesiltas: Bound state solution of the Schrodinger equation for Mie potential. J. Math. Chem. 43, 749 (2007).

[48] C. G. Maitland, M Rigby, E. B. Smith, W. A. Wakeham: Intermolecular Forces. Oxford Univ. Press, Oxford (1987).

[49] I. I. Goldman, V. D. Krivchenkov: Problems in Quantum Mechanics. Pergamon Press, New York(1961)

[50] A. Kratzer: Die ultraroten Rotationsspektren der Halogenwasserstoffe . Z. Phys. 3, 289 (1920).

[51] E. Fues: Das Eiegenschewingungaspekrum zweiatomiger Moleküle in der Undulationsmechanik. Ann. Phys. 80, 367 (1926).

[52] P. Alberto, R. Lisboa, M. Malheiro and A. S. De Castro: Tensor coupling and pseudospin symmetry in nuclei. Phys. Rev. C 71, 034313(2005).

[53] S. Dong, J. G. Ravelo and S.-H. Dong: Analytical approximations to the l-wave solutions of the Schrdinger equation with an exponential-type potential. Phys. Phys. Scr. 76, 393 (2009).

[54] W. C. Qiang and S. H. Dong: Analytical approximations to the solutions oülef the Manning-Rosen potential with centrifugal term. Phys. Lett. A 368, 13 (2007).

[55] C. L. Pekeris: The rotation-Vibration coupling in diatomic molecules. Phys. Rev. 45, 98 (1934).

[56 ] C. S. Jia, T. Chen, L.G. Cui: Approximate analytical solutions of the Dirac equation with the generalized Poschl-Teller potential including the the pseudo-centrifugal term. Phys. Lett. A 373, 1621 (2009)

[57] X.Y. Liu, G. F. Wei, X.W. Cao: Spin Symmetry for Dirac Equation with thetrigonometric Poschl-Teller Potential.Int. J. Ther Phys.49,343 (2010)

Table 1 : The energy levels for the chosen parameters A=2.09 , B=1.58 , m=10, $\alpha = 0.3$, C=10. The energy levels are shown for different values of $\gamma$ and $\kappa$.

| (n, $\ell$ ,j) | (E, $\gamma = 0$) | (E, $\gamma = 2$, $\kappa > 0$) | (n, $\ell$ ,j) | (E, $\gamma = 2, \kappa < 0$) |
|---|---|---|---|---|
| $0s_{1/2}$ | 0.0075 | 0.0150 | | |
| $1s_{1/2}$ | 0.0300 | 0.0900 | | |
| $1p_{1/2}$ | 0.0950 | 0.1250 | $1p_{3/2}$ | 0.0750 |
| $2d_{3/2}$ | 0.2410 | 0.3250 | $2d_{5/2}$ | 0.2150 |
| $2f_{5/2}$ | 0.2950 | 0.3650 | $2f_{7/2}$ | 0.2350 |

Table 2 : the variation of the parameter $\Omega(=\kappa(\kappa+1)+2\kappa\gamma+\gamma(\gamma+1))$ as a function of the coupling constant $\gamma$ for two states.

| $\gamma$ | -20 | -10 | -5 | -3 | 1 | 5 | 7 | 10 | 20 |
|---|---|---|---|---|---|---|---|---|---|
| Energy for $1p_{1/2}$ | 0.609 | 0.250 | 0.131 | 0.094 | 0.125 | 0.205 | 0.256 | 0.345 | 0.789 |
| $Q$ for $1p_{1/2}$ ($\kappa = 1$) | 342 | 72 | 12 | 2 | 6 | 42 | 72 | 132 | 462 |
| $Q$ $1p_{3/2}$ ($\kappa = -2$) | 462 | 132 | 42 | 20 | 2 | 12 | 30 | 72 | 342 |